\documentstyle{article}
\begin{document}
\begin{center}
{\Large \bf General noncommuting curvilinear coordinates and fluid Mechanics.}
\end{center}

\begin{center}
\textbf{S. A. Alavi}

\textit{Department of Physics, Sabzevar university of Tarbiat Moallem, Sabzevar, P. O. Box 397,
 Iran.}\\
\textit{Sabzevar House of Physics, Asrar Laboratories of Physics, Laleh square, Sabzevar, Iran.}\\

\textit{Email: alavi@sttu.ac.ir, alialavi@fastmail.us}
 \end{center}



\emph{ We show that restricting the states of a charged particle to the lowest Landau level introduces noncommutativity between general curvilinear coordinate operators. The cartesian , circular cylindrical and spherical polar coordinates are three special cases of our quite general method. The connection between $U(1)$ gauge fields defined on a general  noncommuting curvilinear coordinates and fluid mechanics is explained. We also recognize the Seiberg-Witten map from general noncommuting to commuting variables as the quantum correspondence of the Lagrange to Euler map in fluid mechanics.}\\

 Recently there have been much interest in the study of physics in noncommutative spaces(NCS)[4-22], not only because the  NCS is necessary when one studies the low energy effective theory of D-brane with B field background, but also because 
 in the very tiny string scale or at very high energy situation, the effects of noncommutativity of space may appear. 
 In the literatures the noncommutative quantum mechanics and noncommutative quantum field theory have been studied extensively and the main approach is based on the Weyl-Moyal correspondence which amounts to replacing the usual product by star product in a noncommutative space. In the usual quantum mechanics (quantum mechanics in commutative space),
  the coordinates and momenta have the following commutation relations :
\begin{equation}
[\hat{x}_{i},\hat{x}_{j}]=0\hspace{1.cm}[\hat{x}_{i},\hat{p}_{j}]=i\delta
_{ij}\hspace{1.cm} [\hat{p}_{i},\hat{p}_{j}]=0 ,
\end{equation}
At very short scales, say string scale, the coordinates may not commute and the commutation relations are as follows : 
\begin{equation}
[\hat{x}_{i},\hat{x}_{j}]=i\theta_{ij}\hspace{1.cm}[\hat{x}_{i},\hat{p}_{j}]=i\delta
_{ij}\hspace{1.cm} [\hat{p}_{i},\hat{p}_{j}]=0 ,
\end{equation}
where $\theta_{ij}$ is an antisymmetric tensor which can be defined as $\theta_{ij}=\frac{1}{2}\epsilon_{ijk}\theta_{k}$.\\

Describing the curvilinear coordinate surfaces by  $q_{1}=constant$, $q_{2}=constant$ and $q_{3}=constant$, we may identify any point by $(q_{1},q_{2},q_{3})$. Let us consider the motion of a particle with charge (e) and mass (m) in the presence of a magnetic field produced by a vector potential $\vec{A}$. The lagrangian is as follows :
\begin{equation}
L=\frac{1}{2}m V^{2}+\frac{e}{c}\vec{A}\cdot\vec{V}-V(x,y)
\end{equation}
where $V_{i}=h_{i}\dot{q}_{i}$(no summation, $i=1,2,3$), are the components of the  velocity of the particle and $h_{i} (i=1,2,3)$ are the scale factors. $V(x,y)$ describes aditional interactions(impurities). In the absence of  $V$  the quantum spectrum consists of the well-known Landau levels. In the strong magnetic field limit only the lowest Landau level is relevant. But the large B limit corresponds to small m, so setting the mass to zero effectively projects onto the lowest Landau level. In the chosen gauge $\vec{A}=(0,h_{1}q_{1}B)$ and in that limit, the Lagrangian(3) takes the following form :
\begin{equation}
L^{\prime}=\frac{e}{c}Bh_{1}h_{2}q_{1}\dot{q}_{2}-V(x,y)
\end{equation}
which is of the form $p\dot{q}-H(p,q)$, and suggests that $\frac{e}{c}Bh_{1}q_{1}$ and $ h_{2}q_{2}$ are canonical conjugates, so we have :
\begin{equation}
[h_{1}q_{1},h_{2}q_{2}]=-i\frac{\hbar c}{eB}
\end{equation}
which can be written in general form :
\begin{equation}
[h_{i}q_{i},h_{j}q_{j}]=i\theta_{ij}
\end{equation}
which is the fundamental space-space noncommutativity relation in a general noncommuting curvilinear coordinates. The cartesian [1], circular cylindrical and spherical polar coordinates are three special cases :
\begin{equation}
[x,y]=-i\frac{\hbar c}{eB}
\end{equation}
\begin{equation}
[\rho ,\rho\phi]=\rho[\rho ,\phi]=-i\frac{\hbar c}{eB}
\end{equation}
\begin{equation}
[r,r\theta]=r[r,\theta]=-i\frac{\hbar c}{eB}
\end{equation}

Consider an infinitesimal coordinate transformation on the general coordinates :
\begin{equation}
\delta q_{i}=-f(q_{i})
\end{equation}
We require that eq.(6) remains unchanged under this transformation, which leads to the following equation :
\begin{equation}
-[h^{i}f^{i}(q),h^{j}q^{j}]-[h^{i}q^{i},h^{j}f^{j}(q)]=0
\end{equation}
 Using eq.(6), this leads to:
\begin{equation}
-\frac{1}{h_{k}}\partial_{k}f^{i}(q)\theta^{kj}-\frac{1}{h_{k}}\partial_{k}f^{j}(q)\theta^{ik}=0
\end{equation}
we define :
\begin{equation}
f^{i}=\theta^{ij}g_{j}
\end{equation}
then eq.(12) becomes :
\begin{equation}
\theta^{i\ell}\nabla_{k}g_{\ell}\theta^{kj}+\theta^{j\ell}\nabla_{k}g_{\ell}\theta^{ik}=0,
\end{equation}
where $\nabla_{i}\equiv\frac{1}{h_{i}}\partial_{i}$. Since $\theta_{ij}$ is nonsingular and antisymmetric, we obtain :
\begin{equation}
\nabla_{k}g_{\ell}-\nabla_{\ell}g_{k}=0
\end{equation}
or :
\begin{equation}
g_{\ell}=\nabla_{\ell}f
\end{equation}
Therefore we have :
\begin{equation}
f^{i}=\theta^{ij}\nabla_{j}f
\end{equation}
 It is easy to show that :
\begin{equation}
\nabla_{i}f^{i}=\nabla\cdot \vec{f}=0
\end{equation}
which means that the transformations are volume preserving. The cartesian [2]case is obtained by setting $h_{i}=1$.
The volume preserving transformations form the link between noncommuting coordinates and fluid mechanics. In fluid mechanics there are two different but physically equivalent pictures of description. The first usually refered as Eulerian, uses as the coordinates the space dependent fields of velocity, density and some thermodynamic variables. The second, Lagrangian description uses the coordinates of the particles $\vec{Q}(q_{i},t)$ labeled by the set of the parameters $q_{i}$, which could be considered as the initial positions $\vec{q}=\vec{Q}(t=0)$ and time $t$. These initial positions $\vec{q}$ as well as the coordinates $\vec{Q}(q_{i},t)$ belong to some domain $D\subseteq R^{3}$, see [3] for more discussions. The continuum description is invariant against volume preserving transformations of $\vec{q}$, and in particular against the specific volume preserving transformations (17), provided the fluid coordinate $Q$ transforms as a scalar :
\begin{equation}
\delta_{f}Q=f^{i}(q)\frac{1}{h_{i}}\frac{\partial}{\partial q_{i}}Q=\theta^{ij}\frac{1}{h_{i}}\partial_{i}Q
\frac{1}{h_{j}}\partial_{j}f
\end{equation}
The connection between Lagrange fluid and noncommuting coordinates is understood from their common invariance against the volume preserving transformation. Eq.(19) can be written in terms of a bracket defined for functions of $q$ by :
\begin{equation}
\left\{O_{1}(q),O_{2}(q) \right\}=\theta^{ij}\frac{1}{h_{i}}\partial_{i}O_{1}(q)\frac{1}{h_{j}}\partial_{j}O_{2}(q)
\end{equation}
It is worth to mention that with this bracket we have :
\begin{equation}
\left\{h_{i}q_{i},h_{j}q_{j} \right\}=\theta_{ij}
\end{equation}
So it is obvious that commutators for a noncommutative field theory can be obtained from calssical commutators by  replacing brackets by $-i$ times commutators. \\
In what follows, it is shown that the noncommuting field theory that emerges from the Lagrange fluid is a noncommuting $U(1)$ gauge theory. To see this, we define the evolving portion of $q$ by :
\begin{equation}
Q^{i}(t,q)=h^{i}q^{i}+\theta^{ij}A_{j}(t,q)
\end{equation}
By the help of this and eq.(19), we have :
\begin{equation}
\delta_{f}A_{i}=\nabla_{i}f+\left\{A_{i},f \right\}
\end{equation}
If we replace the bracket by $(-i)$ times the commutator, the gauge transformation for a noncommuting $U(1)$ gauge potential $\hat{A}_{i}$ is exactly obtained. Furthermore the gauge field $\hat{F}_{ij}$ , emerges from the bracket of two Lagrange coordinates :
\begin{equation}
\left\{Q^{i},Q^{j}\right\}=\theta^{ij}+\theta^{im}\theta^{jn}F_{mn}
\end{equation}
where :
\begin{equation}
\hat{F}_{mn}=\nabla_{m}\hat{A}_{n}-\nabla_{n}\hat{A}_{m}+\left\{\hat{A}_{m},\hat{A}_{n}\right\}
\end{equation}

which is one the fundamental  formula in noncommuting gauge theory. \\
Now we shall discuss the Seiberg-Witten map [4] by the field analogy. The Seiberg-Witten map is a map from ordinary gauge fields $a_{\mu}$ to noncommutative gauge fields $\hat{A}_{\mu}$, which is local to any finite order in $\theta$ and has the following further property. Suppose that two ordinary gauge fields $a_{\mu}$ and $a^{\prime}_{\mu}$ are equivalent by an ordinary gauge transformation by $U=exp(i\lambda)$. Then the corresponding noncommutative gauge field $\hat{A}_{\mu}$ and $\hat{A}^{\prime}_{\mu}$ will also be gauge equivalent by a noncommutative gauge transformation by $U=exp(i\hat{\lambda)}$. However $\hat{\lambda}$ will depend on both $\lambda$ and $A$. If $\hat{\lambda}$ were a function of  $\lambda$  only, the ordinary and noncommutative gauge groups would be the same; since $\hat{\lambda}$ is a function of $A$ as well as $\lambda$, we do not get any well defined mapping between the gauge groups and we get an identification only of the gauge equivalence relations.
Using the Euler formulation of fluid mechanics we obtain an explicit formula for the Seiberg-Witten map. In the Euler description $\vec{Q}$ is promoted to an independent variable and renamed $\vec{r}$. Dynamics is described by the space-time dependent density $\rho(t,\vec{r})$ and velocity $v(t,\vec{r})$. Let $\rho$ and $v$ be the fluid density and velocity at a given point in space-time. They satisfy the continuity equation :
\begin{equation}
\frac{\partial\rho}{\partial t}+\nabla\cdot (\rho \vec{v})=0
\end{equation}
The relation between the Lagrange and Euler description is given by the following formula :
\begin{equation}
\rho(t,\vec{r})=\int dq \delta \left(\vec{Q}(t,\vec{q})-\vec{r}\right)
\end{equation}
\begin{equation}
\rho(t,\vec{r})v(t,\vec{r})=\vec{j}(t,\vec{r})=\int dq \frac{\partial}{\partial t}\vec{Q}(t,\vec{q})\delta \left(\vec{Q}(t,\vec{q})-\vec{r}\right)
\end{equation}

The continuity equation follows from the definition of 4-dimensional flow :
\begin{equation}
j^{\mu}(t,\vec{r})=\int dq \frac{\partial}{\partial t}Q^{\mu}\delta \left(\vec{Q}-\vec{r}\right)
\end{equation}
One can easily show that the right hand side of this equation is invariant under the transformation (19),
and therefore viewed as a function of $\hat{A}$, it is gauge invariant. Because of  continuity equation i.e. eq.(26), we have    $\partial_{\mu}\epsilon_{\alpha\beta\mu}j^{\mu}=0$, so $\epsilon_{\alpha\beta\mu}j^{\mu}$ can be considered as the curl of an Abelian vector:

\begin{equation}
\epsilon_{\alpha\beta\mu}\int dq \frac{\partial}{\partial t}Q^{\mu}\delta(\vec{Q}
-\vec{r})\propto\frac{1}{h_{\alpha}h_{\beta}}\left[\partial_{\alpha} (h_{\beta}a_{\beta})-\partial_{\beta} (h_{\alpha}a_{\alpha})\right]+constant
\end{equation}

(no summation on indices $\alpha$ and $\beta$). This is the(inverse) Seiberg-Witten map, relating $a$ to $\hat{A}$ . The special case of cartesian coordinate[1] is obtained by choosing $h_{i}=1(i=1,2,3)$.\\

In conclusion we have obtained the noncommutativity relations between general curvilinear coordinates operators. Our method is quite general so the cartesian, circular cylindrical and spherical polar coordinates are three special cases of this method. 
This generalization is interesting, because not all physical problems are well adapted to solutions in cartesian coordinates.\\
We have also studied the connection between $U(1)$ gauge fields defined on a general noncommuting curvilinear coordinates and fluid mechanics, especially the the relation between Seiberg-Witten map in noncommutative gauge theory and the Euler map in fluid mechanics.\\ 

\textbf{References. }\\
Because of the huge number of papers on the issue of noncommutativity , hereby we apologize for all the related works which have not been quoted.\\
1. Jackiw R, 2003 Annales. HenriPoincare4S2 S913\\
2. Jackiw R, Pi S.-Y. and Polychronakos 2002 Annals. Phys. 301 157\\
3. Antoniou I and  Pronko G hep-th/0106119\\
As  original papers on noncommutativity see :\\
4. Seiberg N and Witten E 1999 JHEP 9909 032\\
5. Ardalan F, Arfaei H and Sheikh-Jabbari M M 1999 JHEP 9902 016\\
For papers on noncommutative filed theory see for examples :\\
6. Chaichian M, Sheikh-Jabbari M M and Tureanu A 2001 Phys. Rev. Lett. 86 2716\\
7. Pires C A des 2004 J. Phys. G 30 B41\\
8. Correa D H, Fosco C D, Schaposnik F A and Torroba G 2004 JHEP 0409 064\\
9. Khoze V V and Levell J 2004 JHEP 0409 019\\
10. Fatollahi A H and Mohammadzadeh H 2004 Eur. Phys. J. C. 36 113\\
11. Blohmann C 2004 Int. J. Mod. Phys. A 19 5693\\
12. Reichenbach T 2005 Phys. Lett. B 612 275\\
13. Calmet X 2005 Phys. Rev. D 71 085012\\
For papers on noncommutative quantum mechanics see for examples :\\
14. Durhuus B and Jonsson T 2004 JHEP 0410 050\\
15. Zahn J 2004 Phys. Rev. D. 70 107704\\
16. Pinzul A and Stern A 2004 Phys. Lett. B 593 279.\\
17. Kokado A, Okamura T and Saito T 2004 Phys. Rev. D. 69 125007\\
18. Bellucci S and Yeranyan A 2005 Phys. Lett. B 609 418\\
19. Fosco C D and Torroba G 2005 Phys. Rev. D. 71 065012\\
20. Alavi S A 2003 Chin. Phys. Lett. 20 605\\
21. Alavi S A 2005 Mod. Phys. Lett. A. 20 1013\\
22. Alavi S A 2006  Mod. Phys. LettA. 21 4941 (2006)1\\

\end{document}